# Comments on

# "Recent Developments in Plasmon-Assisted Photocatalysis - - a Personal Perspective",

# Cond-Mat ArXiv:200900286, Appl. Phys. Lett. 117, 130501 (2020).

Alexander O. Govorov[1] and Lucas V. Besteiro[2]

[1] Department of Physics and Astronomy, Ohio University, Athens OH 45701, USA

[2] Biomedical Research Centre (CINBIO), Universidade de Vigo, 36310 Vigo, Spain

E-mails: govorov@ohio.edu, lucas.v.besteiro@uvigo.es

**Abstract:** The authors of preprint [1] and paper [2], Y. Sivan and Y. Dubi, made several wrong and inconsistent comments on our papers [3,4]. Moreover, the authors of [1,2] addressed in their comments features that were not present in our paper [4]. In this preprint, we go through the above-mentioned comments made in [1,2], which we found to be wrong and misleading.

**Page 2 in [1]:** "The starting point of our work was a simple question - what happens to a small piece of metal when it is continuously illuminated? This simple looking question turns out to be hard to answer. Specifically, there seems to be a clash between the näive intuitive answer "it heats up" and the strict physical statement that since the metal is out of equilibrium, temperature is no longer well-defined and one cannot talk about heating at all. This may have been the reason that several theoretical papers tried to answer the aforementioned question by considering only how

*light would affect the electron distribution inside the NP (see e.g., Refs. [62 and 63 {Ref. 3 in this commentary}]) or considering heating by taking the temperature of the electrons as a fixed parameter (and guessing it rather than calculating it), and even worse, assuming that phonons do not heat up at all[64] {Ref. 4 in this commentary}. It is, however, not hard to appreciate that while ignoring heat generation and heat leakage to the environment may be valid at the early stages of an ultrashort excitation, these effects cannot be ignored when studying the steady-state case."*

**The above text in [1] includes "… even worse, assuming that phonons do not heat up at all [64]" - this comment in [1] is completely false. Our paper [4] (cited in [1] as [64]) fully includes all necessary thermal effects -** the heating of the lattice and the transfer of electronic energy to the lattice. On page 2762 in [4], we write: "…In the following step, these hot and warm electrons emit phonons and locally increase the lattice temperature." Therefore, in our paper, the lattice temperature $T_{lattice}$ is increased as compared to the ambient temperature $T_0$. The formalism in our paper [4] fully incorporates the phonon-relaxation and heating effects via the relaxation term in the kinetic equation (39) on p. 2766:

$$R_n = \frac{1}{\hbar}\langle n|\hat{\Gamma}\hat{\rho}|n\rangle = \frac{\rho_{nn} - f_F(\varepsilon_n, T_{lattice})}{\tau_{\varepsilon, phonons}} + \ldots ,$$

where the function $f_F(\varepsilon_n, T_{lattice})$ is the Fermi distribution at the lattice temperature. Following the typical experimental situation in the CW illumination regime, optical excitation in our paper [4] is assumed to be weak (i.e. $T_{lattice}$ is close to $T_0$), and, therefore, the optical response functions, such as absorption and the rates of generation of excited electrons, can be evaluated at room temperature, as it is commonly done in textbooks on the optical properties of solids. Our approach is fully justified.

**The phrase in [1]:** " … by taking the temperature of the electrons as a fixed parameter (and guessing it rather than calculating it), …". **The above comment is completely false because our paper [4] does not contain this parameter ("the temperature of the electrons") at all. Therefore, we neither have taken a fixed value for this parameter, nor "guessed" it.** Our paper [4] does not have the parameter "electronic temperature." In [4], we derived our results directly from the quantum master equation and did not employ the electronic-temperature approximation. Our approach is more general as compared to the electronic-T approximation, which is a well-established approximation in literature. From our non-thermal electron distribution in [4], one could easily obtain an equation for the electronic temperature. However, we did not do this since we did not need this parameter to accomplish our goals in [4].

Regarding the above comment "…by taking the temperature of the electrons as a fixed parameter …" in [1,2], this comment seems to relate to a previous misunderstanding of the authors of [1] (given by them in [5]), which we already addressed in a previous response [6].

**Page 2 in [1]:** "Importantly, our calculations have shown clearly that the dominance of thermal effects over non-thermal ("hot" carrier) effects become more significant as the illumination intensity becomes lower. This result invalidates a common claim [64] that since the temperature rise associated with low illumination intensity is small with respect to the ambient temperature, then, "hot" carrier effects are dominant for low intensities."

**The phrase, in [1]:** "This result invalidates a common claim [64] that since the temperature rise associated with low illumination intensity is small with respect to the ambient temperature, then, "hot" carrier effects are dominant for low intensities." **It is a strange and wrong statement about our work [4] – we did not claim the above in [4].** Here, we should make two notes about our view on the hot-electron and photothermal effects:

(1) **We think that the nonthermalized hot-electron contribution at high energies (with the excitation energies $\Delta E = E - E_F \sim 1\,eV$) is the leading (dominating) term to the excited carriers' distribution at weak optical excitation.** The thermalized electrons cannot contribute essentially at high energies assuming the low intensity limit since the Fermi distribution function is exponentially small for high-excitation energies: $f_F(E) \approx e^{-\frac{E-E_F}{kT}}$. As an example, for a photoheating of $\Delta T = 10\,K$, we have $f_F(E) \sim 2.3 \cdot 10^{-17}$, when $T = 303\,K$ and $\Delta E = E - E_F = 1\,eV$. Therefore, in experiments with photo-injection from a metal to a semiconductor, the nonthermalized hot-electrons with high energies will play the major role, i.e. the photo-excited nonthermalized high-energy (hot) electrons will govern photo-physical and photochemical responses at nearly room temperature. This situation with the injection of nonthermalized hot electrons occurs in a large number of photochemical experiments (e.g. in the TiO2-Au systems) and photodetector measurements (using e.g. Au-Si and Ag-Si systems).

(2) **We think that both the plasmonic photothermal effect and the hot-electron generation effect are important and complement each other, and the relative roles of these effects will depend on the particular plasmonic structure and overall system configuration.**

A few comments about our paper [4]. Our paper [4] (cited as [64] in [1]) concerns mainly the hot-electron generation phenomenon. In particular, in our paper [4], we show that, in nanocrystals with typical sizes ( ~ 10nm or larger), the number of non-thermalized high-energy (hot) electrons (with energies $\sim \hbar\omega \sim 1eV$) **is smaller** than the numbers of photo-generated non-thermalized electrons with low energies. In our paper [4], we study the electronic structure of the plasmon and focus only on the non-thermalized electrons. By computing the rates of excitation of non-thermalized electrons with low and high energies, we obtain the energy efficiency of generation of high-energy (hot) electrons. The latter is an important parameter for hot-electron photochemistry and photodetectors.

**We add, that, in [4], we are not interested in fully thermalized electrons given by the Fermi distribution, which are trivial to compute**. In our formalism [4], the thermalized electrons are kept as a background inside the function $f_F(\varepsilon_n, T_{lattice})$. It is trivial to find the number of thermalized electrons for a given parameter $T_{lattice}$ from the function $f_F(\varepsilon_n, T_{lattice})$.

**Page 6 in [1]:** "Additional common misconceptions (such as the incorrect claim on the dominance of non-thermal effects over thermal effects at low illumination intensity [64], the absence of transverse temperature uniformities, the dependence of the number of "hot" electrons on particle size and shape (see correction of [64] in [5]) …"

The above phrase "such as the incorrect claim on the dominance of non-thermal effects over thermal effects at low illumination intensity [64]" **is another straight falsehood about our paper, since we do not made this claim in [4].** Again, we repeat that, in our opinion, both the plasmonic photothermal effect and the hot-electron generation effect are significant in plasmonic nanocrystals

(also see our detailed comments above). Moreover, one of us is the author of some of the early papers on the plasmonic photothermal effects in nanostructures – being, for example, among the first researchers who described the collective heating mechanism [8] – a mechanism commonly used now in many experiments. In our paper [4], we studied the generation of high-energy (hot) electrons and the quantum electronic structure of the plasmon. Simultaneously, we were not interested in the fully thermalized electrons, that are expected and trivial to compute. Our focus was on the efficiency of generation of high-energy (hot) electrons, that is a useful parameter for different emerging applications.

**In the comment in [1]:** "Additional common misconceptions … the dependence of the number of "hot" electrons on particle size and shape (see correction of [64] in [5])…" **Again, it is a misleading and ill-defined comment.** Regarding the previous publications by Y. Dubi and Y. Sivan [5,7], we commented on these papers in our preprint [6]. We have shown in [6] that the statements given in the papers [5,7], which are related to our work [4], are misleading and wrong. To add, we note that the generation of hot electrons is, of course, a size- and shape-dependent effect since it originates from optical absorption processes, which in plasmonic systems strongly depend on the shape and size of a nanocrystal.

**To finally conclude, the authors of [1,2] made strange and wrong statements about our study [4]. We are under the impression that the authors of [1,2] do not understand the content and terminology of our paper [4].**